\documentclass[a4paper]{spie}  
\usepackage{amsmath,amsfonts,amssymb}

\usepackage[utf8]{inputenc}   
\usepackage[english]{babel}   
\usepackage[usenames,dvipsnames]{xcolor} 
\usepackage[colorlinks=true, allcolors=blue, pdftex]{hyperref}
\usepackage{aas_macros}
\usepackage{paralist}
\usepackage{graphicx}

\title{The SOXS Instrument Control Software \\ approaching the PAE }

\author[*,a]{Davide Ricci}
\author[a]{Bernardo Salasnich}
\author[a]{Andrea Baruffolo}

\author[l]{Jani Achrén}
\author[b]{Matteo Aliverti}
\author[v]{José A. Araiza-Durán}
\author[n]{Iair Arcavi}
\author[b]{Laura Asquini}
\author[a]{Federico Battaini}
\author[g]{Sagi Ben-Ami}
\author[g]{Alex Bichkovsky}
\author[v]{Anna Brucalassi}
\author[g]{Rachel Bruch}
\author[b]{Lorenzo Cabona}
\author[b]{Sergio Campana}
\author[c]{Giulio Capasso}
\author[a]{Enrico Cappellaro}
\author[a]{Riccardo Claudi}
\author[c]{Mirko Colapietro}
\author[e]{Rosario Cosentino}
\author[f]{Francesco D'Alessio}
\author[b]{Paolo D'Avanzo}
\author[c]{Sergio D'Orsi}
\author[c]{Massimo Della Valle}
\author[k]{Rosario Di Benedetto}
\author[a]{Simone Di Filippo}
\author[g]{Avishay Gal-Yam}
\author[b]{Matteo Genoni}
\author[e]{Marcos Hernandez Díaz}
\author[g]{Ofir Hershko}
\author[j,q]{Jari Kotilainen}
\author[j,q]{Hanindyo Kuncarayakti}
\author[b]{Marco Landoni}
\author[r]{Gianluca Li Causi}
\author[c]{Laurent Marty}
\author[q]{Seppo Mattila}
\author[k]{Matteo Munari}
\author[b]{Luca Oggioni}
\author[e]{Hector Pérez Ventura}
\author[b]{Giorgio Pariani}
\author[m]{Giuliano Pignata}
\author[a]{Kalyan Radhakrishnan}
\author[s]{Stephen Smartt}
\author[s]{Michael Rappaport}
\author[b]{Marco Riva}
\author[h]{Adam Rubin}
\author[c]{Salvatore Savarese}
\author[c]{Pietro Schipani}
\author[w,k]{Salvatore Scuderi}
\author[u]{Maximilian Stritzinger}
\author[f]{Fabrizio Vitali}
\author[s]{David Young}
\author[k]{Ricardo Zanmar Sanchez}

\affil[a]{INAF -- Osservatorio Astronomico di Padova, Vicolo dell’Osservatorio 5, I-35122, Padua, Italy }
\affil[b]{INAF -- Osservatorio Astronomico di Brera, Via Bianchi 46, I-23807, Merate, Italy }
\affil[c]{INAF -- Osservatorio Astronomico di Capodimonte, Sal. Moiariello 16, I-80131, Naples, Italy }
\affil[e]{FGG-INAF, TNG, Rambla J.A. Fernández Pérez 7, E-38712 Breña Baja (TF), Spain }
\affil[f]{INAF -- Osservatorio Astronomico di Roma, Via Frascati 33, I-00078 M. Porzio Catone, Italy }
\affil[g]{Weizmann Institute of Science, Herzl St 234, Rehovot, 7610001, Israel }
\affil[h]{ESO, Karl Schwarzschild Strasse 2, D-85748, Garching bei München, Germany }
\affil[i]{Max-Planck-Institut für Extraterrestrische Physik, Giessenbachstr. 1, D-85748 Garching, Germany }
\affil[j]{Finnish Centre for Astronomy with ESO (FINCA), FI-20014 University of Turku, Finland}
\affil[k]{INAF -- Osservatorio Astrofisico di Catania, Via S. Sofia 78 30, I-95123 Catania, Italy }
\affil[l]{Incident Angle Oy, Capsiankatu 4 A 29, FI-20320 Turku, Finland }
\affil[m]{Instituto de Alta Investigaci\'on, Universidad de Tarapac\'a, Arica, Casilla 7D, Chile}
\affil[n]{The School of Physics and Astronomy, Tel Aviv University, Tel Aviv 69978, Israel}
\affil[o]{Dark Cosmology Centre, Juliane Maries Vej 30, DK-2100 Copenhagen, Denmark }
\affil[p]{Aboa Space Research Oy, Tierankatu 4B, FI-20520 Turku, Finland}
\affil[q]{Tuorla Observatory, Dept. of Physics and Astronomy, FI-20014 University of Turku, Finland }
\affil[r]{INAF - Istituto di Astrofisica e Planetologia Spaziali, Rome, Italy}
\affil[s]{Astrophysics Research Centre, Queen's University Belfast, Belfast, BT7 1NN, UK }
\affil[u]{Aarhus University, Ny Munkegade 120, D-8000 Aarhus, Denmark }
\affil[v]{INAF-Osservatorio Astrofisico di Arcetri}
\affil[w]{INAF -- Istituto di Astrofisica Spaziale e Fisica Cosmica, Milan, Via Corti 12, I-20133 Milano, Italy}

\authorinfo{$^*$Contact information:
  D.R: davide.ricci@inaf.it, +39-049-829-3480
}

\begin{document}
\maketitle

\begin{abstract}
  The Instrument Control Software of SOXS (Son Of X-Shooter), the
  forthcoming spectrograph for the ESO New Technology Telescope at the
  La Silla Observatory, has reached a mature state of development and
  is approaching the crucial Preliminary Acceptance in Europe
  phase. Now that all the subsystems have been integrated in the
  laboratories of the Padova Astronomical Observatory, the team
  operates for testing purposes with the whole instrument at both
  engineering and scientific level. These activities will make use of
  a set of software peculiarities that will be discussed in this
  contribution. In particular, we focus on the synoptic panel, the
  co-rotator system special device, on the Active Flexure Compensation
  system which controls two separate piezo tip-tilt devices.
\end{abstract}


\section{Introduction}
\label{sec:intro}

SOXS (\textit{``Son Of X-Shooter''}), the new spectrograph for visible
and infrared wavelengths for the NTT (New Technology Telescope)
mainly focused on transients, is approaching the Preliminary
Acceptance in Europe, thanks to the progresses of the Assembly,
Integration and Test phase of its several subsystems.
\cite{2022SPIE12189E..1IY, 2022SPIE12189E..0LL, 2022SPIE12189E..0AA,
  2022SPIE12188E..44S, 2022SPIE12187E..0CG, 2022SPIE12184E..84C,
  2022SPIE12184E..83A, 2022SPIE12184E..82R, 2022SPIE12184E..81A,
  2022SPIE12184E..80B, 2022SPIE12184E..7ZV, 2022SPIE12184E..5IC,
  2022SPIE12184E..0OS}

SOXS INS (Instrument Control Software) is developed under the VLTSW
(Very Large Telescope Software) standard environment. 
The most of its components such as motors and detectors are
then natively supported as so-called ``standard devices'', and have
been set up via configuration files.
However, there are devices which are not natively supported by the
VLTSW. For these to be interfaced with the high-level software is
necessary to develop a Function Block software at PLC level, and a VLTSW
``special device'' driver.

As a part of a team effort presented in this conference,
\cite{soxs-ricci, soxs-genoni, soxs-schipani, soxs-araiza-duran,
  soxs-radha, soxs-asquini, soxs-colapietro, soxs-cosentino,
  soxs-vitali, soxs-claudi, soxs-scaudo, soxs-battaini}
we discuss a brief overview of SOXS, exploiting its
Synoptic panel (see Sect.~\ref{sec:syn}).
Then, we focus on two of the INS special devices:
the co-rotator system (see Sect.~\ref{sec:crot}), and
the piezo tip-tilt AFCs for Active Flexure Compensation (see
Sect.~\ref{sec:afc}),
adding details to our previous proceedings\cite{2020SPIE11452E..2QR,
  2018SPIE10707E..1GR}.
Conclusions are presented in Sect.~\ref{sec:conc}.

\section{Synoptic panel}
\label{sec:syn}

\begin{figure} [h]
  \centering
    \includegraphics[width=0.7\textwidth]{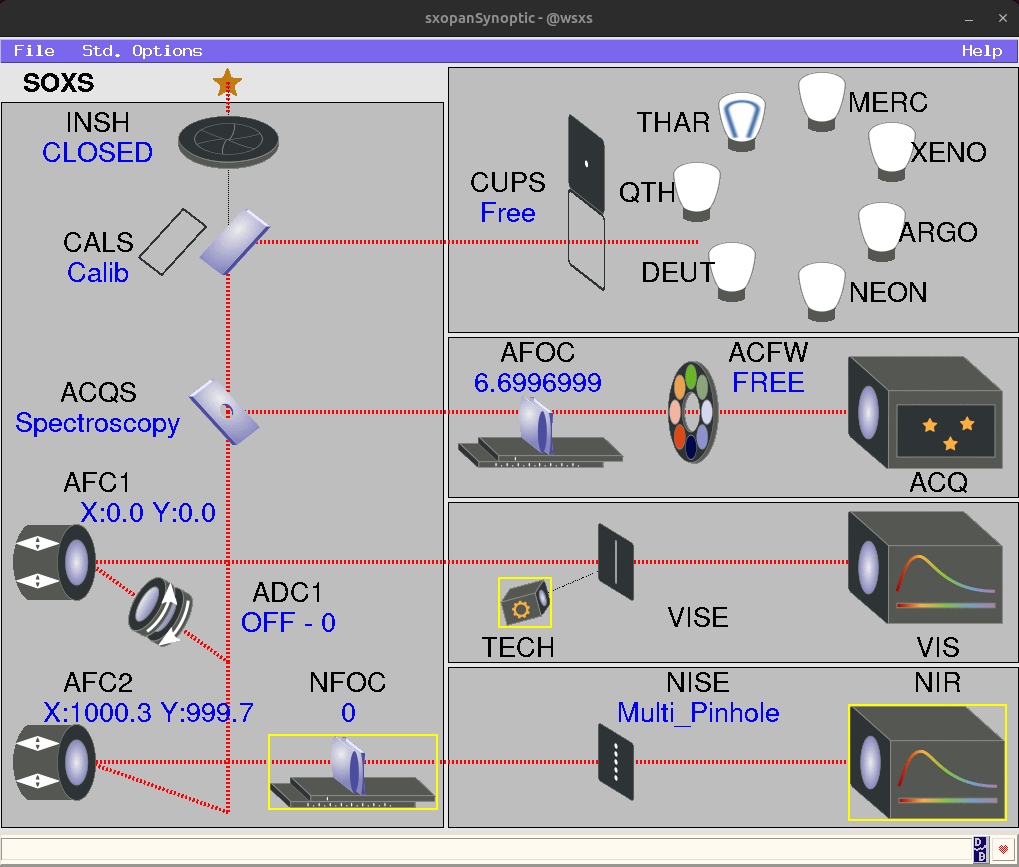}
  \caption[CROT]
  { \label{fig:syn} SOXS Synoptic panel.}
\end{figure}

The Synoptic panel (see Fig.~\ref{fig:syn}) is a custom developed GUI,
showing the current configuration of the instrument in a graphical
way. This panel will display the light path and the current position
for movable functions.

In the figure is visible the entrance shutter \textsf{INSH}, used to
let the telescope beam in or make the instrument light-tight when
performing calibrations using internal light sources.
Then, \textsf{CALS} linear motor allows to select the instrument input
source: either the lamps of the Calibration Unit, provided with
\textsf{CUPS}, a dedicated motor to insert a Pinhole or the light
coming from the target field on sky.
\textsf{ACQS} can direct all the beam to the Acquisition Camera
\textsf{ACQ} and its filter wheel \textsf{ACFW} and refocuser
\textsf{AFOC} for Imaging observation, or only the external part for
secondary guiding, while the rest of the beam feeds the spectrographs.
\textsf{AFC}s Tip-Tilt Mirrors compensate for mechanical flexures due
to derotation, while \textsf{ADC} corrects the Atmospheric Dispersion
and \textsf{NFOC} adjusts the focus for the near-infrared
spectrograph.
\textsf{VISE} and \textsf{NISE} linear motors are the slit exchangers
of the \textsf{VIS} and \textsf{NIR} spectrograph, respectively.

ESO recently requested the Synoptic panel in its more recent
instruments in order to have a quick look of its instruments.  In
general, synoptic panels follow the opto-mechanical design of the
instrument, where components and light path are reproduced to mock
their ``real'' position.
For SOXS, we decided to change approach towards a ``subway map''
concept, to improve the operator's readability.
In particular, most of the gif icons animate depending on their
status (shutter \texttt{OPEN} or \texttt{CLOSE}, calibration unit
lamps \texttt{ON} or \texttt{OFF}, type of the slits, position of the
mirrors.)
Finally, simulated devices are bordered to improve awareness.
Given the positive feedback by the team, we decided to broaden this
approach also for the VLT ERIS instrument and for the LBT SHARK-NIR
coronagraph.

\section{Co-rotator}
\label{sec:crot}

\begin{figure} [h]
  \centering
    \includegraphics[width=0.7\textwidth]{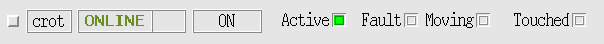}
  \caption[CROT]
  { \label{fig:crot}
    CROT custom widget in the SOXS ICS panel.}
\end{figure}

SOXS will be held by one of the NTT Nasmyth foci derotator. Then,
during operations, the rotation of the whole instrument would also
affect the cabling of the several devices.
For that reason, a co-rotator system has been set up, and will
compensate for the de-rotator activity.
Two linear potentiometers
have been mounted on an interface flange between the instrument and
the co-rotator, and check the relative rotation between them. Their
signals give the feedback to the servo motor driver about the relative
displacement between the two systems in the rotation.
The potentiometers adapter accepts these signals as inputs and returns
their difference as output.
A safety switch
is not connected, and
is inserted in the telescope interlock chain in order to disable also
the rotator when the co-rotator is in \texttt{OFFLINE} state.
In case the switch is reached, the interlock signal is raised and the
final stage that gives power to the servo motor is disabled; then, a
manual repositioning of the co-rotator is needed in order to disable
the interlock and reactivate the final stage.
When the final stage is disabled and the co-rotator is not working,
the \texttt{Active} output signal is low.  The driver is controlled via
digital and analog I/O:
\begin{itemize}
\item two analog inputs will give information to the driver about the
  relative rotation between the instrument and the co-rotator;
\item a digital input, coming from a safety switch, will stop
  the motor in case of emergency / safety risks;
\item two digital input signals, coming from the Beckhoff PLC, will
  control the activation and reset of the driver: \texttt{Enable} (Enable the
  power stage); \texttt{Fault Reset} (Reset error message);
\item two digital output signals will interface with the Beckhoff PLC
  giving information about the state of the driver: \texttt{Active}
  (Indication of the Operating State); \texttt{No Fault} (Indication
  of the Fault condition).
\end{itemize}
A custom widget in the ICS panel (see Fig.~\ref{fig:crot}) allows to
change the state of the special device, and to monitor the above flags.

\section{Active Flexure Compensation}
\label{sec:afc}

\begin{figure}[h]
  \centering
    \includegraphics[width=0.7\textwidth]{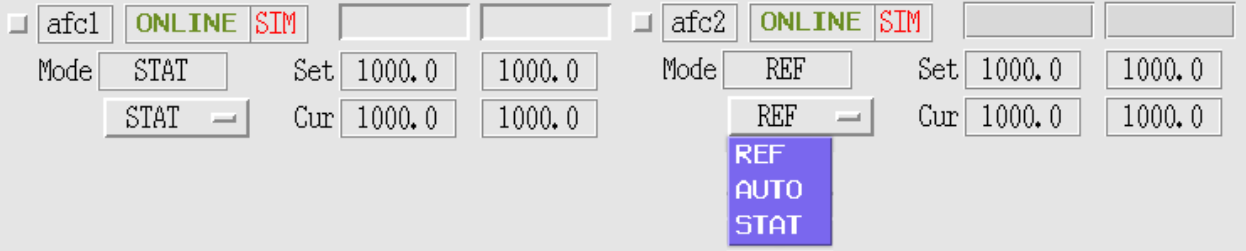}
  \caption[AFC]
  { \label{fig:afc}
    AFC custom widgets in the SOXS ICS panel.}
\end{figure}

When rotating, SOXS will be subjected to flexures, potentially
displacing the target in relation to \textsf{VISE} and
\textsf{NISE}. To address this issue, two piezo-actuated tip-tilt
mirrors \textsf{AFC1} and \textsf{AFC2} are positioned to correct for
these flexures.

They are controlled by the Instrument Control System (INS) through the
instrument Programmable Logic Controller (PLC) via analog signals,
with one signal per axis. Since these devices are not standard VLT
actuators, a custom ``special device'' has been developed to
automatically compensate the position at a $1 \rm Hz$ frequency
(\texttt{AUTO} mode).  The loop is controlled by the Instrument
Workstation and it is calculated with a look-up table based on the
rotation position, interpolating X and Y values which have been
measured during the AIV phase with a step of 5 degrees.
The \textsf{AFC1}, responsible for the visible arm, will also correct
for ADC "wobbling" if necessary, taking the ADC prism angle into
account.
Other operation modes of the \textsf{AFC}s include the \texttt{STAT} Mode,
in which the piezos are kept at a fixed position set via a
\texttt{SETUP} command; and the \texttt{REF} Mode, in which they are
placed at a predefined fixed position required for system alignment.

Two custom widgets in the ICS panel (see Fig.~\ref{fig:afc}) allows to
change the operation mode and eventually select a custom position.


\section{Conclusion}
\label{sec:conc}

We presented several peculiarities in the development of the SOXS
Instrument Control Software, focusing on the Synoptic panel for the
quick look of the instrument status and setup, and on two non standard
devices: the co-rotator that compensate the derotation of the flange,
and the whole connected instrument; and the Active Flexure
Compensation system that corrects the trajectory of the light path on
the spectrograph to compensate for the effects introduced by the derotation.
Further tests are ongoing in order to face the Preliminary Acceptance
in Europe phase, looking forward for SOXS first light in La Silla.


\bibliographystyle{spiebib} 
\bibliography{ricci-soxs} 

\begin{thebibliography}{10}

  
\bibitem{2022SPIE12189E..1IY}
{Young}, D.~R., {Landoni}, M., {Smartt}, S.~J., {Campana}, S., {D'Avanzo}, P.,
  and others
 ``{The Son-Of-X-Shooter
  (SOXS) data-reduction pipeline},'' in [{\em Software and Cyberinfrastructure
  for Astronomy VII}{\nolinebreak\hspace{0.1em}]},  {\em Society of
  Photo-Optical Instrumentation Engineers (SPIE) Conference Series} {\bf
  12189},  121891I (Aug. 2022).

\bibitem{2022SPIE12189E..0LL}
{Landoni}, M., {Marty}, L., {Young}, D., {Asquini}, L., {Smartt}, S.~J.,
  and others
 ``{The quality check system architecture for Son-Of-X-Shooter
  SOXS},'' in [{\em Software and Cyberinfrastructure for Astronomy
  VII}{\nolinebreak\hspace{0.1em}]},  {\em Society of Photo-Optical
  Instrumentation Engineers (SPIE) Conference Series} {\bf 12189},  121890L
  (Aug. 2022).

\bibitem{2022SPIE12189E..0AA}
{Asquini}, L., {Landoni}, M., {Young}, D., {Marty}, L., {Smartt}, S.~J.,
  and others
  ``{Dynamic scheduling for SOXS instrument: environment, algorithms and
  development},'' in [{\em Software and Cyberinfrastructure for Astronomy
  VII}{\nolinebreak\hspace{0.1em}]},  {\em Society of Photo-Optical
  Instrumentation Engineers (SPIE) Conference Series} {\bf 12189},  121890A
  (Aug. 2022).

\bibitem{2022SPIE12188E..44S}
{Scuderi}, S., {Bellassai}, G., {Di Benedetto}, R., {Martinetti}, E.,
  and others
 ``{The vacuum and
  cryogenics system of the SOXS spectrograph},'' in [{\em Advances in Optical
  and Mechanical Technologies for Telescopes and
  Instrumentation}{\nolinebreak\hspace{0.1em}]},  {\em Society of Photo-Optical
  Instrumentation Engineers (SPIE) Conference Series} {\bf 12188},  1218844
  (Aug. 2022).

\bibitem{2022SPIE12187E..0CG}
{Genoni}, M., {Scaudo}, A., {Li Causi}, G., {Cabona}, L., {Landoni}, M.,
  and others
 ``{Progress on
  the simulation tools for the SOXS spectrograph: exposure time calculator and
  end-to-end simulator},'' in [{\em Modeling, Systems Engineering, and Project
  Management for Astronomy X}{\nolinebreak\hspace{0.1em}]},  {Angeli}, G.~Z.
  and {Dierickx}, P., eds., {\em Society of Photo-Optical Instrumentation
  Engineers (SPIE) Conference Series} {\bf 12187},  121870C (Aug. 2022).

\bibitem{2022SPIE12184E..84C}
{Claudi}, R., {Radhakrishnan}, K., {Battaini}, F., {Campana}, S., {Schipani},
  and others
 ``{SOXS AIT: a paradigm for system
  engineering of a medium class telescope instrument},'' in [{\em Ground-based
  and Airborne Instrumentation for Astronomy IX}{\nolinebreak\hspace{0.1em}]},
  {Evans}, C.~J., {Bryant}, J.~J., and {Motohara}, K., eds., {\em Society of
  Photo-Optical Instrumentation Engineers (SPIE) Conference Series} {\bf
  12184},  1218484 (Aug. 2022).

\bibitem{2022SPIE12184E..83A}
{Araiza-Dur{\'a}n}, J.~A., {Pignata}, G., {Brucalassi}, A., {Battaini}, F.,
  and others
 ``{The integration and alignment phase for the
  acquisition and guiding system of SOXS},'' in [{\em Ground-based and Airborne
  Instrumentation for Astronomy IX}{\nolinebreak\hspace{0.1em}]},  {Evans},
  C.~J., {Bryant}, J.~J., and {Motohara}, K., eds., {\em Society of
  Photo-Optical Instrumentation Engineers (SPIE) Conference Series} {\bf
  12184},  1218483 (Aug. 2022).

\bibitem{2022SPIE12184E..82R}
{Radhakrishnan Santhakumari}, K.~K., {Battaini}, F., {Claudi}, R., {Slemer},
  and others
 ``{From assembly to the
  complete integration and verification of the SOXS common path},'' in [{\em
  Ground-based and Airborne Instrumentation for Astronomy
  IX}{\nolinebreak\hspace{0.1em}]},  {Evans}, C.~J., {Bryant}, J.~J., and
  {Motohara}, K., eds., {\em Society of Photo-Optical Instrumentation Engineers
  (SPIE) Conference Series} {\bf 12184},  1218482 (Aug. 2022).

\bibitem{2022SPIE12184E..81A}
{Aliverti}, M., {Battaini}, F., {Radhakrishnan}, K., {Genoni}, M., {Pariani},
  and others
  ``{SOXS mechanical integration and verification in Italy},'' in [{\em
  Ground-based and Airborne Instrumentation for Astronomy
  IX}{\nolinebreak\hspace{0.1em}]},  {Evans}, C.~J., {Bryant}, J.~J., and
  {Motohara}, K., eds., {\em Society of Photo-Optical Instrumentation Engineers
  (SPIE) Conference Series} {\bf 12184},  1218481 (Aug. 2022).

\bibitem{2022SPIE12184E..80B}
{Battaini}, F., {Radhakrishnan}, K., {Claudi}, R., {Munari}, M., {S{\'a}nchez},
  and others
 ``{The
  internal alignment and validation of a powered ADC for SOXS},'' in [{\em
  Ground-based and Airborne Instrumentation for Astronomy
  IX}{\nolinebreak\hspace{0.1em}]},  {Evans}, C.~J., {Bryant}, J.~J., and
  {Motohara}, K., eds., {\em Society of Photo-Optical Instrumentation Engineers
  (SPIE) Conference Series} {\bf 12184},  1218480 (Aug. 2022).

\bibitem{2022SPIE12184E..7ZV}
{Vitali}, F., {Aliverti}, M., {D'Alessio}, F., {Genoni}, M., {Scuderi}, S.,
  and others
  ``{Progress on the SOXS NIR spectrograph AIT},'' in [{\em Ground-based and
  Airborne Instrumentation for Astronomy IX}{\nolinebreak\hspace{0.1em}]},
  {Evans}, C.~J., {Bryant}, J.~J., and {Motohara}, K., eds., {\em Society of
  Photo-Optical Instrumentation Engineers (SPIE) Conference Series} {\bf
  12184},  121847Z (Aug. 2022).

\bibitem{2022SPIE12184E..5IC}
{Cosentino}, R., {Hernandez}, M., {Ventura}, H., {Campana}, S., {Claudi}, R.,
  and others
 ``{Laboratory test of the VIS
  detector system of SOXS for the ESO-NTT Telescope},'' in [{\em Ground-based
  and Airborne Instrumentation for Astronomy IX}{\nolinebreak\hspace{0.1em}]},
  {Evans}, C.~J., {Bryant}, J.~J., and {Motohara}, K., eds., {\em Society of
  Photo-Optical Instrumentation Engineers (SPIE) Conference Series} {\bf
  12184},  121845I (Aug. 2022).

\bibitem{2022SPIE12184E..0OS}
{Schipani}, P., {Campana}, S., {Claudi}, R., {Aliverti}, M., {Baruffolo}, A.,
  and others
 ``{Progress on the SOXS
  transients chaser for the ESO-NTT},'' in [{\em Ground-based and Airborne
  Instrumentation for Astronomy IX}{\nolinebreak\hspace{0.1em}]},  {Evans},
  C.~J., {Bryant}, J.~J., and {Motohara}, K., eds., {\em Society of
  Photo-Optical Instrumentation Engineers (SPIE) Conference Series} {\bf
  12184},  121840O (Aug. 2022).

\bibitem{soxs-ricci}
Ricci, D. et~al., ``{The SOXS Instrument Control Software challenges
  approaching the PAE},'' in [{\em {Software and Cyberinfrastructure for
  Astronomy VIII}}{\nolinebreak\hspace{0.1em}]},  {\em \procspie} {\bf
  13101-91} (2024).

\bibitem{soxs-genoni}
Genoni, M., ``{SOXS NIR: optomechanical integration and alignment, optical
  performance verification before full instrument assembly},'' in [{\em
  {Ground-based and Airborne Instrumentation for Astronomy
  X}}{\nolinebreak\hspace{0.1em}]},  {\em \procspie} {\bf 13096-104} (2024).

\bibitem{soxs-schipani}
Schipani, P., ``{Walking with SOXS towards the transient sky},'' in [{\em
  {Ground-based and Airborne Instrumentation for Astronomy
  X}}{\nolinebreak\hspace{0.1em}]},  {\em \procspie} {\bf 13096-65} (2024).

\bibitem{soxs-araiza-duran}
Araiza-Durán, J.~A. et~al., ``{Final alignment and image quality test for the
  acquisition and guiding system of SOXS},'' in [{\em {Ground-based and
  Airborne Instrumentation for Astronomy X}}{\nolinebreak\hspace{0.1em}]},
  {\em \procspie} {\bf 13096-267} (2024).

\bibitem{soxs-radha}
{Radhakrishnan Santhakumari}, K.~K. et~al., ``{Integration and verification of
  the SOXS Instrument},'' in [{\em {Ground-based and Airborne Instrumentation
  for Astronomy X}}{\nolinebreak\hspace{0.1em}]},  {\em \procspie} {\bf
  13096-268} (2024).

\bibitem{soxs-asquini}
Asquini, L., ``{Automated scheduler for the SOXS instrument: design and
  performance},'' in [{\em {Software and Cyberinfrastructure for Astronomy
  VIII}}{\nolinebreak\hspace{0.1em}]},  {\em \procspie} {\bf 13101-90} (2024).

\bibitem{soxs-colapietro}
Colapietro, M., ``{The integration of the SOXS control electronics towards the
  PAE},'' in [{\em {Ground-based and Airborne Instrumentation for Astronomy
  X}}{\nolinebreak\hspace{0.1em}]},  {\em \procspie} {\bf 13096-107} (2024).

\bibitem{soxs-cosentino}
Cosentino, R., ``{Characterisation and assessment of the SOXS Spectrograph
  UV-VIS Detector System},'' in [{\em {Ground-based and Airborne
  Instrumentation for Astronomy X}}{\nolinebreak\hspace{0.1em}]},  {\em
  \procspie} {\bf 13096-105} (2024).

\bibitem{soxs-vitali}
Vitali, F., ``{The status of the NIR arm of the SOXS Instrument toward the
  PAE},'' in [{\em {Ground-based and Airborne Instrumentation for Astronomy
  X}}{\nolinebreak\hspace{0.1em}]},  {\em \procspie} {\bf 13096-106} (2024).

\bibitem{soxs-claudi}
Claudi, R.~U., ``{SOXS System engineering from design to installation.
  Challenges and results},'' in [{\em {Modeling, Systems Engineering, and
  Project Management for Astronomy XI}}{\nolinebreak\hspace{0.1em}]},  {\em
  \procspie} {\bf 13099-63} (2024).

\bibitem{soxs-scaudo}
Scaudo, A., ``{End-to-End simulation framework for astronomical spectrographs:
  SOXS, CUBES and ANDES},'' in [{\em {Modeling, Systems Engineering, and
  Project Management for Astronomy XI}}{\nolinebreak\hspace{0.1em}]},  {\em
  \procspie} {\bf 13099-4} (2024).

\bibitem{soxs-battaini}
Battaini, F., ``{A new tool to quickly and precisely align an optical bench in
  combination with a pCMM},'' in [{\em {Advances in Optical and Mechanical
  Technologies for Telescopes and Instrumentation
  VI}}{\nolinebreak\hspace{0.1em}]},  {\em \procspie} {\bf 13100-32} (2024).

\bibitem{2020SPIE11452E..2QR}
  {Ricci}, D., {Baruffolo}, A., {Salasnich}, B., {De Pascale}, M., {Campana}, S.,
  and others
  ``{Development status of the SOXS instrument control software},'' in [{\em
    Software and Cyberinfrastructure for Astronomy
    VI}{\nolinebreak\hspace{0.1em}]},  {Guzman}, J.~C. and {Ibsen}, J., eds.,
  {\em Society of Photo-Optical Instrumentation Engineers (SPIE) Conference
    Series} {\bf 11452},  114522Q (Dec. 2020).

\bibitem{2018SPIE10707E..1GR}
{Ricci}, D., {Baruffolo}, A., {Salasnich}, B., {Fantinel}, D., {Urrutia}, J.,
  and others
 ``{Architecture of the SOXS instrument control software},''
  in [{\em Software and Cyberinfrastructure for Astronomy
  V}{\nolinebreak\hspace{0.1em}]},  {\em Society of Photo-Optical
  Instrumentation Engineers (SPIE) Conference Series} {\bf 10707},  107071G
  (July 2018).

\bibitem{2014SPIE.9152E..07K}
{Kiekebusch}, M.~J., {Lucuix}, C., {Erm}, T.~M., {Chiozzi}, G., et~al., ``{PC
  based PLCs and ethernet based fieldbus: the new standard platform for future
  VLT instrument control},'' in [{\em Software and Cyberinfrastructure for
  Astronomy III}{\nolinebreak\hspace{0.1em}]},  {\em \procspie} {\bf 9152},
  915207 (July 2014).

\bibitem{2014SPIE.9152E..0ID}
{Duhoux}, P., {Knudstrup}, J., {Lilley}, P., {Di Marcantonio}, P., {Cirami},
  R., and {Mannetta}, M., ``{VLT instruments: industrial solutions for
  non-scientific detector systems},'' in [{\em Software and Cyberinfrastructure
  for Astronomy III}{\nolinebreak\hspace{0.1em}]},  {\em \procspie} {\bf 9152},
   91520I (July 2014).

\end{thebibliography}

\end{document}